\documentclass{ws-ijmpb}
\usepackage{graphicx}
\begin{document}

\markboth{B. B. Baizakov, A. Bouketir, A. Messikh, A. Benseghir,
B. A. Umarov} {Variational analysis of flat-top solitons in
Bose-Einstein condensates}

\title{VARIATIONAL ANALYSIS OF FLAT-TOP SOLITONS \\ IN BOSE-EINSTEIN CONDENSATES}

\author{B. B. BAIZAKOV}
\address{Physical-Technical Institute, Uzbek Academy of Sciences, Tashkent, 100084, Uzbekistan\\
baizakov@uzsci.net}

\author{A. BOUKETIR}
\address{Department of Mathematical Sciences Community College, King Fahd University of \\ Petroleum
and Minerals, Dhahran, 1261, Saudi Arabia\\
abouketir@kfupm.edu.sa}

\author{A. MESSIKH}
\address{Department of Computer Science, Faculty of ICT,
International Islamic University Malaysia, Kuala Lumpur, P.O. Box
10, 50728, Malaysia\\
amessikh@yahoo.com}

\author{A. BENSEGHIR, \ B. A. UMAROV}
\address{Department of Computational and Theoretical Sciences, Faculty of
Science, \\ International Islamic University Malaysia, Kuantan,
P.O. Box 141, 25710, Malaysia \\
bakhram@iiu.edu.my}

\maketitle

\begin{history}
\received{Day Month Year} \revised{Day Month Year}
\end{history}

\begin{abstract}
Static and dynamic properties of matter-wave solitons in dense
Bose-Einstein condensates, where three-body interactions play a
significant role, have been studied by a variational approximation
(VA) and numerical simulations. For experimentally relevant
parameters, matter-wave solitons may acquire a flat-top shape,
which suggests employing a super-Gaussian trial function for VA.
Comparison of the soliton profiles, predicted by VA and those
found from numerical solution of the governing Gross-Pitaevskii
equation shows good agreement, thereby validating the proposed
approach.
\end{abstract}


\section{Introduction}

The existence and properties of solitons in Bose-Einstein
condensates (BEC) has been the subject of considerable interest
over the recent decade (for a review see articles
\cite{soliton-reviews} and books
\cite{malomed-book,kevrekidis-book}). All of the main types of
matter-wave solitons, such as dark \cite{dark}, bright
\cite{bright}, and gap \cite{gap}, have been observed in the
experiments. Dark solitons emerge in a BEC with repulsive
interactions between atoms (nonlinearity is defocusing), while for
the existence of bright solitons the interatomic interaction has
to be attractive (nonlinearity is focusing). Gap solitons develop
in repulsive condensates loaded in periodic potential of the
optical lattice. While in homogeneous condensates solitons exist
due to the balance between nonlinearity and dispersion, in a
repulsive BEC subject to a periodic potential of the optical
lattice, gap solitons come out of the interplay between
nonlinearity and periodicity of the medium.

Nowadays the conditions for the existence of these main types of
solitons are usually created in magnetic and optical traps for
cold quantum gases~\cite{bongs}. Meantime, the experimental
techniques for trapping and manipulation of BEC's are progressing
and novel conditions for the localized matter-waves are emerging.
In this regard, development of atom chips for BEC's
\cite{atom-chip} and further advances in Feshbach resonance
management of atomic interactions \cite{feshbach} can be mentioned
as two important examples. Specifically, the atom chip technology
combined with techniques to suppress three-body recombination
\cite{search} allows to produce long-lived condensates with
increased density, where the contribution of three-body scattering
is dominant. On the other hand, even in condensates with normal
density, inhibition of the two-body s-wave scattering length by a
Feshbach resonance technique gives rise to enhanced role of the
three-body effects. In these conditions the governing mean-field
Gross-Pitaevskii equation (GPE) for the dynamics of BEC includes
higher order nonlinear terms, alongside with the usual cubic term.
As a consequence, localized matter-waves exhibit novel features
both in their shape and dynamic behavior, compared to matter-wave
solitons of the conventional GPE.

In this work we focus on a specific class of bright solitons, so
called ``flat-top" solitons, which remain less explored in the
context of BEC. Flat-top solitons in BEC's emerge when the
repulsion between atoms, originating from three-atom collisions,
prevails the attraction resulting from two-atom interactions. In
terms of the governing GPE, this implies that the defocusing
quintic nonlinearity is stronger than the focusing cubic
nonlinearity. In the experiments such a situation is realized when
the density of the condensate is high, for instance in BEC's on
atom chips, or when the usual two-body interactions are suppressed
by means of a Feshbach resonance technique. The properties BEC
described by GPE with cubic-quintic nonlinearity and generic trap
potential can be explored using the Lagrangian formalism. Our
objective is the development of a Lagrangian based variational
approximation (VA) for flat-top solitons using the super-Gaussian
trial function. Static version of the VA provides stationary shape
of the soliton, while the time-dependent version can be used for
studying small amplitude oscillations around stationary states.

The paper is structured as follows. In section II we formulate the
model and present the governing equations. In sections III and IV,
respectively, the static and dynamic versions of the VA have been
developed. In concluding section V we formulate our main findings.

\section{The model and main equations}

The dynamics of a quasi-1D Bose-Einstein condensate with two- and
three-body interactions, confined to an external trap potential,
is described by the following Gross-Pitaevskii equation (see e.g.
\cite{adv,zhang})
\begin{equation}\label{gpe0}
i\hbar\frac{\partial \psi}{\partial t} = -
\frac{\hbar^2}{2m}\frac{\partial^2 \psi}{\partial x^2} + V(x)\psi
+ \frac{g_1 N}{2\pi a_{\bot}^2}|\psi|^2\psi + \frac{g_2
N^2}{3\pi^2a_{\bot}^4}|\psi|^4\psi,
\end{equation}
where $\psi(x,t)$ is the mean-field wave function of the
condensate, $m$ is the atomic mass, $g_1=4\pi\hbar^2a_s/m$ is the
coefficient of cubic nonlinearity with $a_s$ being the atomic
s-wave scattering length, characterizing the strength of two-body
interactions, $g_2$ is the coefficient of quintic nonlinearity
responsible for three-body interactions, $a_{\bot} =
\sqrt{\hbar/m\omega_{\bot}}$ is the transverse harmonic oscillator
length, $N$ is the number of atoms comprising the BEC, and $V(x)=m
\omega_x^2 x^2/2$ is the harmonic trap potential in the axial
direction. Strengths of the trap potential in the axial and radial
directions are given, respectively, through the trap frequencies
$\omega_x$ and $\omega_{\bot}$.

In deriving the Eq. (\ref{gpe0}) from the original 3D GPE it is
assumed that the transverse confinement of the condensate is
strong enough, so that its radial degrees of freedom are fixed. In
this condition characteristic energies of the transverse
excitations $\hbar \omega_{\bot}$ are much greater than the energy
from the nonlinear term $E_{int}=4\pi \hbar^2 a_s n_0/m$, where
$a_s, n_0, m$ are the atomic $s$-wave scattering length, peak
density of the condensate and the atomic mass, respectively.
Therefore, the dynamic evolution of the condensate is possible
only in the axial direction, while in the radial direction it
remains in the ground state of the strong parabolic trap
(transverse dynamics is frozen). Alternatively, the criterion for
the one-dimensionality of BEC can be expressed as $\sqrt{\hbar/m
\omega_{\bot}} >> \xi = (4\pi n_0 a_s)^{-1/2}$, which can be
understood as the radial harmonic oscillator length being much
greater than the healing length (for $^{87}$Rb: $a_s = 5.6 \, nm$,
$n_0 = 10^{14}$ cm$^{-3}$, $m = 1.45 \times 10^{-25} kg$, $\xi =
0.4 \, \mu m$).

The wave function is subject to the normalization condition,
\begin{equation}
\int_{-\infty }^{+\infty }\left\vert \psi \left( x\right)
\right\vert ^{2}dx=1.  \label{norm1}
\end{equation}
For convenience we reduce the Eq. (\ref{gpe0}) into dimensionless
form by adopting new variables $t \rightarrow \omega_{\bot} t$, $x
\rightarrow x/a_{\bot}$, $\alpha=-g_1 N/ (2\pi a_{\bot}^3 \,
\hbar\omega_{\bot})$, \\ $\beta=g_2 N^2/(3\pi^2 a_{\bot}^6 \,
\hbar\omega_{\bot})$, $U(x)=-V(x)/(\hbar \omega_{\bot})$, $\psi
\rightarrow \sqrt{a_{\bot}}\,\psi$
\begin{equation}
i\psi _{t}+\frac{1}{2}\psi _{xx} + U(x) \psi + \alpha |\psi
|^{2}\psi  -\beta |\psi |^{4}\psi  =0.  \label{gpe1}
\end{equation}
It is instructive to estimate the experimentally realistic values
for the main parameters $\alpha$ and $\beta$ in Eq. (\ref{gpe1}).
According to data of \cite{zhang} for $^{87}$Rb condensate $g_1
\simeq 5~\hbar~\times~10^{-11}$~cm$^3$/s, $g_2 \simeq
4~\hbar~\times~10^{-26}$~cm$^6$/s, i.e. both nonlinear terms in
Eq. (\ref{gpe0}) are positive (repulsive). However, in our case
for the existence of self-trapped localized matter-waves we assume
that the coefficient of cubic nonlinearity $g_1 = 4\pi \hbar^2
a_s/m$ is shifted to negative (attractive) domain via change of
the s-wave scattering length $a_s$ by a Feshbach resonance
technique. The strength of transverse confinement used in
matter-wave soliton experiments~\cite{bright} $\omega_{\bot}=700
\div 800 \, s^{-1}$ provides $a_{\bot} \sim 1 \, \mu$m. Then, for
the number of atoms in $^{87}$~Rb BEC $N \sim 10^4$ we get the
following values for the dimensionless coefficients of cubic and
quintic nonlinear terms $\alpha=-g_1 N/ (2\pi a_{\bot}^3 \,
\hbar\omega_{\bot}) = 2 (|a_s|/a_{\bot})\,N \sim 100$, $\beta =
q_2 N^2/(3 \pi^2 a_{\bot}^6 \hbar \omega_{\bot}) \sim 200 $. In
subsequent numerical simulations we employ the above estimates for
the coefficients.

Exact one soliton solutions of Eq. (\ref{gpe1}) in the absence of
external potential ($U(x)~=~0$) were found in Ref.
\cite{pushkarov}. For the case of self-focusing cubic ($\alpha
> 0$) and defocusing quintic ($\beta >0$) nonlinearities, under
normalization condition (\ref{norm1}), the solution is
\begin{equation} \psi (x,t)=\sqrt{\frac{3\alpha}{4\beta}} \
\frac{{\rm tanh}(\eta) \ \exp[i(qx-\mu t)]}{\sqrt{1 +
\mathrm{sech}(\eta) \mathrm{cosh}(x/a)}}, \quad \eta \equiv
\sqrt{\frac{2\beta}{3}}, \quad a \equiv
\frac{1}{\alpha}\frac{\eta}{\mathrm{tanh}(\eta)}, \label{exact1}
\end{equation}
where $q$ and $\mu $ stand for the wave vector and chemical
potential. In presence of a trap potential the Eq. (\ref{gpe1})
does not have analytic solution, and therefore one has to recourse
to approximate methods.

It is appropriate to mention, that cubic-quintic nonlinear
Schr\"odinger equation (\ref{gpe1}) with flat-top solutions
appears in a variety of physical contexts in nonlinear optics
\cite{biswas,akhmediev,gagnon}, fluid dynamics \cite{grimshow},
plasma physics \cite{zhou} and BEC~\cite{as2005}. Recently it was
considered as a model equation describing bright solitons in the
Tonks-Girargeau gas with dipolar interactions~\cite{tg}.
Therefore, the variational approach developed in this paper is of
general interest for the above mentioned fields.

\section{Static variational approximation}

The variational approximation represents one of the important
theoretical tools for investigation of solitons in non-integrable
models \cite{va}. The success of VA essentially depends on the
proper choice of a trial function. In particular, significant
progress has been made with application of VA to nonlinear
Schr\"odinger (NLS) type equations with cubic nonlinearity,
employing Gaussian and hyperbolic secant trial functions. However,
when the higher order nonlinear terms are included in the NLS
equation, the shapes of localized states may significantly deviate
from the above mentioned functions, and one has to consider other
options. The possibility to perform analytic calculations is the
major issue in selection of a trial function.

For the NLS with competing cubic and quintic nonlinearities, when
the soliton features a flat-top shape, a
super-Gaussian~\cite{dimitrevski} and
super-secant~\cite{jovanoski} trial functions were shown to be
appropriate for the description of self-trapping of laser beams in
two-dimensional (2D) cubic-quintic nonlinear media. The behavior
of soliton solution of the NLS equation with arbitrary
nonlinearity near the blow-up point was investigated in
\cite{cooper} by means of VA based on a super-Gaussian trial
function, and accurate estimate for the critical blow-up mass was
found. Among other successful applications of the super-Gaussian
trial function, description of the pulsating localized solutions
of the cubic-quintic complex Ginzburg-Landau equation~\cite{tsoy}
and stationary solutions to the NLS equation in a parabolic-index
fiber~\cite{karlsson}, can be mentioned.

Although the Eq. (\ref{gpe1}) without external potential
($U(x)=0$) has exact one soliton solution (\ref{exact1}), in Ref.
\refcite{deangelis} a variational approximation using the Gaussian
trial function was developed for this case. Justification for the
construction of VA when the exact soliton solution is available,
can be found in some advantages provided by the VA in the analysis
of existence and stability of solitons. For instance, when both of
the nonlinear terms in Eq. (\ref{gpe1}) are focusing, the
localized wave undergoes collapse (unlimited shrinking) if the
norm of the wave function exceeds some critical value. The VA with
a Gaussian trial function accurately predicts the corresponding
threshold norm (the solution ceases to exist at this value of the
norm) \cite{deangelis}. Furthermore, simple analytic relations
between parameters of the localized state provided by VA
facilitates its stability analysis by means of the Vakhitov -
Kolokolov criterion \cite{vk}.

However, the Gaussian trial function restricts the validity of the
developed VA to specific areas of the parameter space of Eq.
(\ref{gpe1}), since it is adequate only if both nonlinear terms
are focusing, or when the effect of repulsive quintic term is weak
compared to the attractive cubic one. In the opposite situation
the localized solution acquires a flat-top shape, and one has to
consider a different trial function. With this motivation in mind
below we develop the VA for Eq. (\ref{gpe1}) using a
super-Gaussian trial function, and apply it for the analysis of
static and dynamic properties of flat-top matter-wave solitons.

\subsection{Stationary wave profile in a free space}

It is instructive to start with considering the variational
solution of Eq.~(\ref{gpe1}) in a free space and comparing the
obtained wave profile with the available exact one soliton
solution (\ref{exact1}). The aim here is twofold. Form one side we
can thereby check the accuracy of the VA, and from the other side,
useful relations between soliton parameters will be obtained.

In absence of external trap potential the governing equation has
the following form
\begin{equation} \label{locgpe}
i\psi_t + \frac{1}{2}\psi_{xx} + \alpha |\psi|^2\psi - \beta
|\psi|^4 \psi  =0.
\end{equation}
Although the coefficient of cubic nonlinearity $\alpha$ can be
rescaled to one by transformations $\psi \rightarrow \sqrt{\alpha}
\psi$ and $\beta \rightarrow \beta/\alpha^2$, we retain it for
convenience of the time dependent VA to be considered later. The
stationary states, which are looked for as $\psi(x,t)=\phi(x)
\exp(-i\mu t)$, satisfy the following equation
\begin{equation}
\mu \phi + \frac{1}{2}\phi_{xx} + \alpha \phi^3 - \beta \phi^5 =
0.
\end{equation}
The Lagrangian density generating this equation is
\begin{equation}
{\cal L} = \frac{1}{4} \phi_x^2 - \frac{\mu}{2}\phi^2 -
\frac{\alpha}{4}\phi^4 + \frac{\beta}{6}\phi^6.
\end{equation}

Since the typical localized solutions of Eq. (\ref{locgpe}) for
competing nonlinearities (attractive cubic and repulsive quintic)
are the "flat-top" solitons, we employ a super-Gaussian trial
function
\begin{equation}\label{ansatz}
\phi(x) = A \exp\left[-\frac{1}{2}\left(\frac{x}{a}\right)^{2
m}\right],
\end{equation}
with $A, a, m$ being the variational parameters, corresponding to
the amplitude, width and super-Gaussian index of the soliton,
respectively.

\begin{figure}
\centerline{\includegraphics[width=8cm, height=6cm]{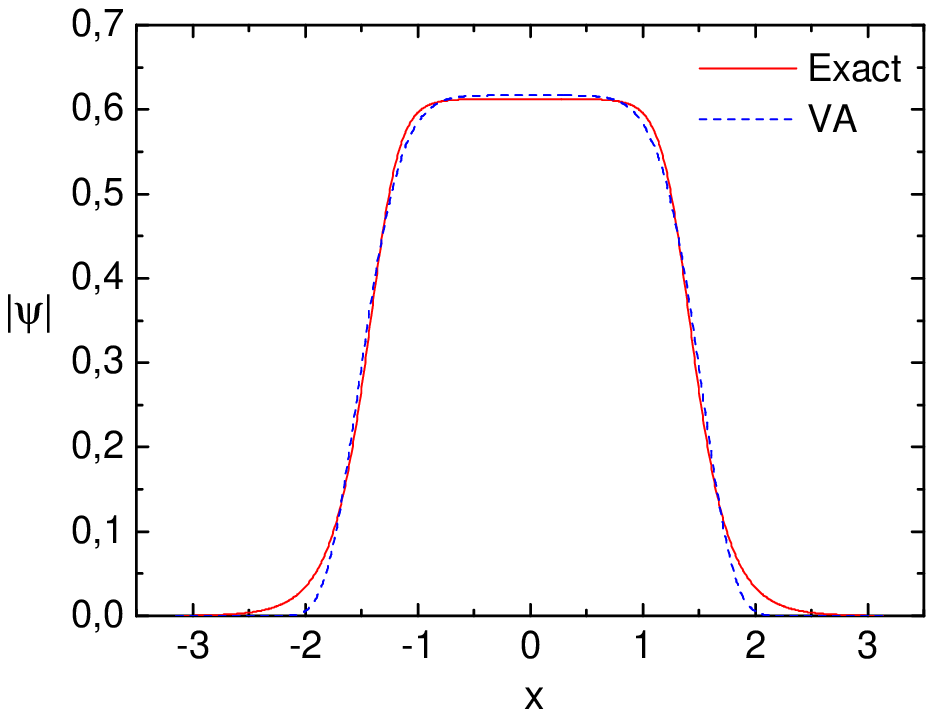}}
\caption{(Color online) Comparison of the exact solution (red
continuous line) given by Eq. (\ref{exact1}) with the prediction
of VA (blue dashed line) for the flat-top soliton according to
Eqs. (\ref{A})-(\ref{M}) for $\alpha=100$, $\beta=200$. Parameters
of the flat-top soliton found from VA are $A=0.62$, $a=1.41$,
$M=0.155$, $\mu=-9.35$, $m=1/2M=3.22$.} \label{fig1}
\end{figure}

The averaged Lagrangian $L=\int_{-\infty}^{\infty} {\cal L} dx$
computed with this ansatz is
\begin{equation}
L = \frac{A^2}{8aM}\Gamma(2-M) - \left(A^2a\mu +
\frac{A^4a\alpha}{2^{M+1}} - \frac{A^6a\beta}{3^{M+1}}
\right)\Gamma(1+M),
\end{equation}
where $\Gamma(x)$ is the gamma function, and the notation $M=1/2m$
is introduced. The stationary values for the amplitude ($A$),
width ($a$) and reduced super-Gaussian index ($M$) of the wave
profile are found from the variational equations
\begin{equation}\label{vaeq}
\frac{\partial L}{\partial A^2} = 0, \quad \frac{\partial
L}{\partial a} = 0, \quad \frac{\partial L}{\partial M} = 0.
\end{equation}
Straightforward calculations, with taking into regard the above
mentioned normalization
\begin{equation} \label{norm}
{\cal N}=\int_{-\infty}^{+\infty} \phi^2(x)dx = 2A^2a \Gamma(1+M)
\equiv 1,
\end{equation}
yield the following relations between variational parameters
\begin{eqnarray}
A &=& \left[\frac{\alpha \ (3/2)^{M+1}}{2\beta + 3^{M+1}\
\Gamma(M)\ \Gamma(2-M)} \right]^{1/2}, \label{A} \\
a &=& \frac{1}{2A^2\Gamma(1+M)}, \label{width} \\
\mu &=& -\frac{3\alpha A^2}{2^{M+2}} + \frac{2\beta A^4}{3^{M+1}},
\label{mu}
\end{eqnarray}
\begin{equation} \label{M}
\Gamma(M)\Gamma(2-M)\left[M^{-1}+\psi(2-M)-\psi(1+M)-
2\,\mathrm{ln}2\right]+\beta
\frac{2\,\mathrm{ln}(3/4)}{3^{M+1}}=0,
\end{equation}
where $\psi(x) = \frac{d}{dx} {\rm ln}\Gamma(x)$ is the digamma
function \cite{abramowitz} (not to be confused with the wave
function). The Eqs. (\ref{A})-(\ref{M}) are sufficient to
determine the four parameters of the localized state ($A, a, m,
\mu$). An example of stationary solution of Eq. (\ref{locgpe}) for
a particular set of parameters is presented in Fig.~\ref{fig1}.
As can be seen from this figure the agreement between the exact
and VA wave profiles for the flat-top soliton is quite good.

An important observation following from the above analysis is
that, in the flat-top regime the super-Gaussian index $M$ (and
therefore $m=1/2M$) for the soliton does not depend on the
coefficient of cubic nonlinearity $\alpha$. In fact this is the
manifestation of above mentioned rescaling property of Eq.
(\ref{locgpe}). This property later will be used in construction
of the time-dependent VA. In the next subsection we extend the
static VA for the case when the flat-top soliton is confined to a
harmonic trap.

\subsection{Flat-top soliton in harmonic trap}

Usual experimental settings for BEC involve magnetic or optical
traps designed for holding and manipulation with matter-waves. In
a confining trap potential the soliton experiences deformation to
some extent with respect to the free space condition. Below we
consider the harmonic trap potential $U(x)=\varepsilon \, x^2$ in
Eq. (\ref{gpe1}) and derive corresponding VA equations.

\begin{figure}
\centerline{\includegraphics[width=8cm, height=6cm]{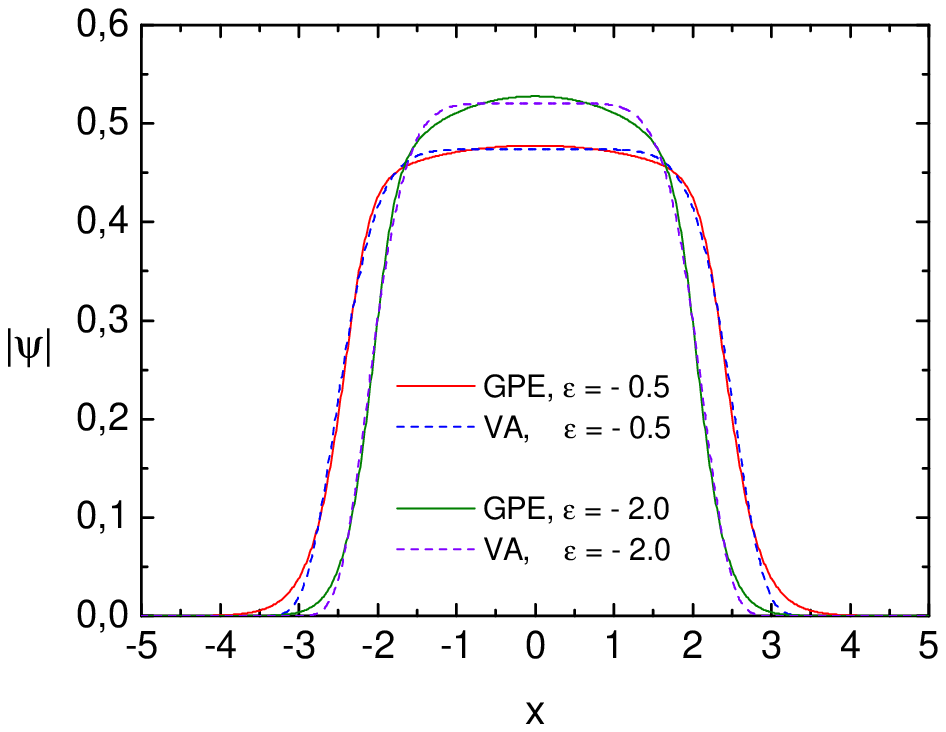}}
\caption{(Color online) Wave profiles of flat-top solitons for two
strengths of the trap potential $\varepsilon = -0.5$ and
$\varepsilon = -2.0$. Solid lines correspond to numerical solution
of the original GPE (\ref{gpe1}), dashed lines are prediction of
the VA. Parameters: $\alpha=100$, $\beta=400$.} \label{fig2}
\end{figure}

\begin{figure}
\centerline{\includegraphics[width=6cm,height=6cm]{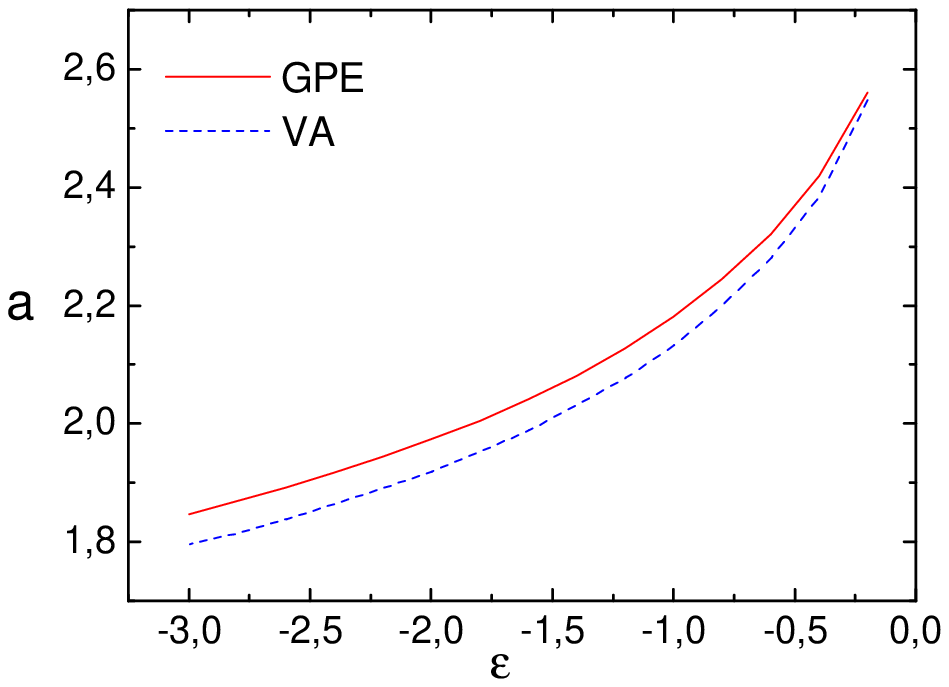}\quad
            \includegraphics[width=6cm,height=6cm]{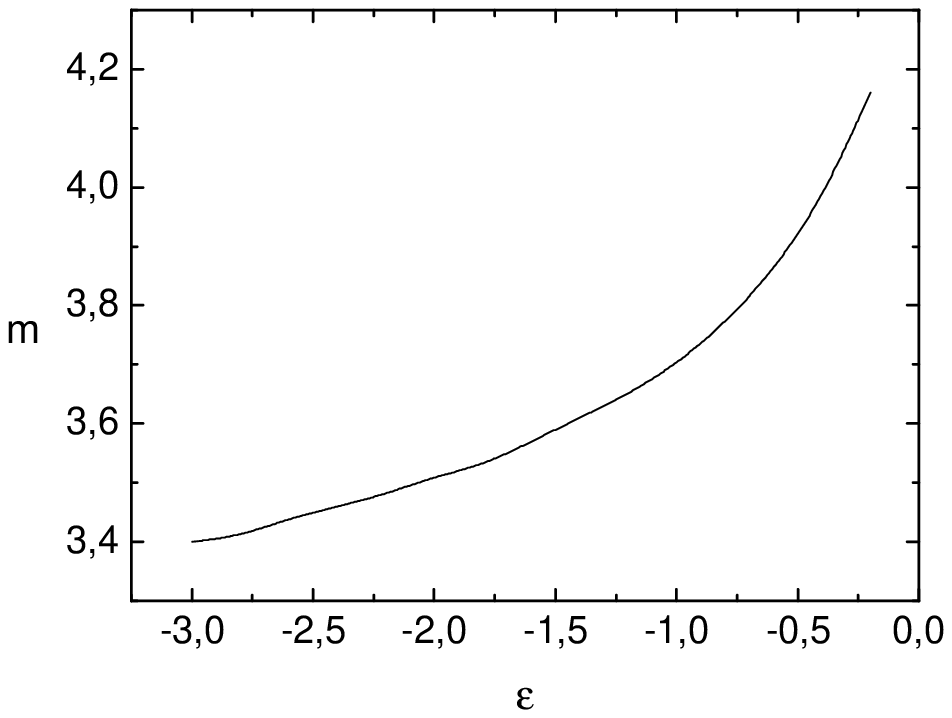}}
\caption{(Color online) Left panel: The width of the flat-top
soliton as a function of the strength of the trap potential. Red
solid line corresponds to numerical solution of the original GPE
(\ref{gpe1}), while blue dashed line is the prediction of VA for
$\alpha = 100$ and $\beta=400$. Right panel: The super-Gaussian
index in Eq. (\ref{ansatz}) as a function of the strength of the
trap potential.} \label{fig3}
\end{figure}

The Lagrangian density generating the Eq. (\ref{gpe1}) is
\begin{equation}
{\cal L} = \frac{1}{4} \phi_x^2 - \frac{\mu}{2}\phi^2 -
\frac{1}{2}U(x)\,\phi^2 - \frac{\alpha}{4}\phi^4 +
\frac{\beta}{6}\phi^6,
\end{equation}
and the corresponding averaged Lagrangian computed with the
super-Gaussian ansatz (\ref{ansatz}) has the form
\begin{equation}
L = \frac{A^2 \, \Gamma(2-M)}{8aM} - \frac{A^2 a^3 \varepsilon \,
\Gamma(1+3M)}{3} - \left(A^2a\mu + \frac{A^4a\alpha}{2^{M+1}} -
\frac{A^6a\beta}{3^{M+1}} \right)\Gamma(1+M).
\end{equation}
Application of the conditions (\ref{vaeq}) to this averaged
Lagrangian yields the following relations between parameters of
the flat-top soliton
\begin{equation}\label{mu1}
\mu= -\frac{2 a^2 \varepsilon
\Gamma(1+3M)}{3\Gamma(1+M)}-\frac{3\alpha A^2}{2^{M+2}} +
\frac{2\beta A^4}{3^{M+1}}.
\end{equation}
The effective width of the soliton is found as the root of the
algebraic equation
\begin{equation}\label{a}
a^4 + p \,a + q = 0, \\
\end{equation}
with the coefficients
$$ p = -\frac{3\alpha}{2^{M+3}\Gamma(1+3M) \, \varepsilon }, \quad q
= \frac{2\beta + 3^{M+1}\Gamma(M)\Gamma(2-M)}{8 \cdot 3^M
\Gamma(1+M) \Gamma(1+3M) \, \varepsilon}. $$ Real and positive
root of the Eq. (\ref{a}) is
\begin{equation}\label{aroot}
a=\frac{1}{2}\sqrt{\frac{2p}{\sqrt{s}} - s} - \frac{\sqrt{s}}{2},
\end{equation}
where
$$s=\frac{4q}{3r}+r, \qquad r=\left(\frac{p^2}{2}+\frac{1}{2}\sqrt{p^4-\frac{256}{27} \
q^3} \ \right)^{1/3}.$$ The counterpart of the Eq. (\ref{M}) for
the trapped soliton has the form
\begin{eqnarray}
\frac{a^4 \, \varepsilon \,
\Gamma(1+3M)\,(2\psi(1+M)-3\psi(1+3M))}{3} + \frac{a\, \alpha \,
(2 \,
\mathrm{ln}2 + \psi(1+M))}{2^{M+3}} &-& \nonumber \\
\frac{\Gamma(2-M) \, (1+M\psi(2-M))}{8 M^2} - \frac{\beta \,
(\mathrm{ln} 3 + \psi(1+M))}{4 \cdot 3^{M+1} \Gamma(1+M)} = 0.
\label{mroot}
\end{eqnarray}
In order to find the shape of the flat-top soliton confined to a
parabolic trap at first we solve this equation with respect to
$M$, substituting $a$ from Eq. (\ref{aroot}). Subsequently the
width $a$ is computed from Eq. (\ref{aroot}) using the value of
$M$ found as a root of Eq. (\ref{mroot}). Next, the amplitude $A$
is computed from the expression for the norm (\ref{norm}), and
chemical potential is found from Eq.~(\ref{mu1}). In the left
panel of Fig.~{\ref{fig2}} we illustrate the shapes of flat-top
solitons for two strengths of the trap potential $\varepsilon$ as
predicted by VA, and as found from the original GPE~(\ref{gpe1})
by imaginary time relaxation method~\cite{chiofalo}.
As can be seen from this figure, stronger parabolic trap leads to
more deformation of the soliton shape. Specifically, under the
effect of a parabolic trap the flat-top soliton shrinks, while its
amplitude increases. Dependence of the width ($a$) and
super-Gaussian index ($m$) as a function of the strength of the
trap potential ($\varepsilon$) is depicted in the right panel of
Fig. \ref{fig3}. The deviation between the results of numerical
solution of the original GPE (\ref{gpe1}) in imaginary time
\cite{chiofalo} and prediction of the VA increases as the trap
potential becomes stronger. However, the discrepancy for the given
range of $\varepsilon$ is less than 3 \%, thus the agreement is
quite good.

\section{Time dependent variational approximation}

The time dependent VA allows to investigate soliton dynamics under
external or internal perturbations, such as varying strength of
the trap potential, chirp imprinting or alternating coefficient of
nonlinear interaction. Below we develop the dynamic version of VA
for the latter case, when the coefficient of cubic nonlinearity in
Eq. (\ref{gpe0}) is periodically varied in time, via the s-wave
scattering length, by means of a Feshbach resonance
technique~\cite{feshbach}. This setting is frequently called as
the nonlinearity management~\cite{malomed-book}.

For simplicity we consider the case when the trap potential is
absent. In dimensionless units the governing equation has the form
\begin{equation} \label{gpe-td}
i\psi_t + \frac{1}{2}\psi_{xx} + \alpha(t) |\psi|^2\psi - \beta
|\psi|^4 \psi  =0,
\end{equation}
where $\alpha(t) = - 2 a_s(t) N/a_{\bot}$ is the time dependent
strength of the cubic nonlinearity. In the following we consider
the periodic $\alpha(t)=\alpha_0 [1+\delta \sin(\omega t)]$
variation of this parameter, with the magnitude $\delta$ and
frequency $\omega$, around its stationary value $\alpha_0$.

To construct the time dependent VA we consider the following
Lagrangian density, generating the Eq. (\ref{gpe-td})
\begin{equation} \label{lagden1}
{\cal L} = \frac{i}{2}(\psi \psi^{\ast}_t - \psi^{\ast}\psi_t) +
\frac{1}{2} |\psi_x|^2 - \frac{\alpha(t)}{2}|\psi|^4 +
\frac{\beta}{3}|\psi(x,t)|^6.
\end{equation}
While selecting the trial function we recall that for the flat-top
soliton the super-Gaussian index $m$ does not explicitly depend on
the strength of cubic nonlinearity $\alpha$, according to Eq.
(\ref{M}). Therefore, when the parameter $\alpha$ is subject to
variations, the following trail function with constant $m$ is
appropriate
\begin{equation}\label{ansatz1}
\psi(x,t) = A(t)
\exp\left[-\frac{1}{2}\left(\frac{x}{a(t)}\right)^{2 m} + i b(t)
x^2 + i \phi(t) \right],
\end{equation}
where $A(t), a(t), b(t)$ and $\phi(t)$ are the time dependent
variational parameters, denoting the amplitude, width, chirp and
phase of the flat-top soliton, respectively.

Substituting the ansatz (\ref{ansatz1}) into Eq. (\ref{lagden1})
and performing spatial integration yields the following averaged
Lagrangian
\begin{eqnarray}
L &=& (a^2 b_t + 2 a^2 b^2) \, \frac{\Gamma(1+3M)}{3 \Gamma(1+M)}
+ \frac{\Gamma(2-M)}{8 M \Gamma(1+M) \, a^2 } +
\phi_t  \nonumber \\
 & & - \frac{\alpha (t)}{2^{M+2} \Gamma(1+M) \, a } + \frac{\beta}{4 \cdot
3^{M+1} \Gamma^2(1+M)\, a^2 },
\end{eqnarray}
where we have used the normalization condition (\ref{norm}) to
eliminate $A$, and employed the notation $M=1/2m$. Variation of
this Lagrangian with respect to time dependent parameters $a$, $b$
and $\phi$ leads to the following equation for the width of the
flat-top soliton

\begin{equation} \label{att}
a_{tt} =\frac{f_1(M)+\beta f_2(M)}{a^3} - \frac{\alpha (t)
f_3(M)}{a^2},
\end{equation}
where
\begin{eqnarray}
f_1(M) &=& \frac{3\Gamma(2-M)}{4 M \, \Gamma(1+3M)}, \quad
f_2(M)=\frac{1}{2\cdot 3^M \, \Gamma(1+M)\Gamma(1+3M)}, \nonumber \\
f_3(M) &=& \frac{3}{2^{M+2}\, \Gamma(1+3M)}  \nonumber.
\end{eqnarray}

\begin{figure}
\centerline{\includegraphics[width=6cm,height=6cm]{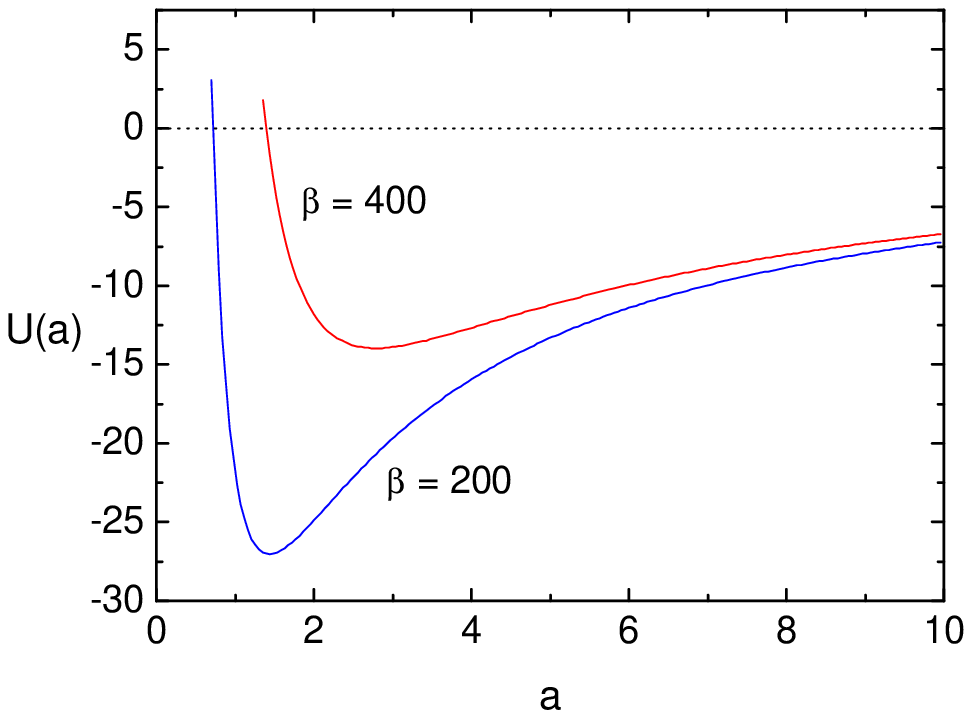}\qquad
            \includegraphics[width=6cm,height=6cm]{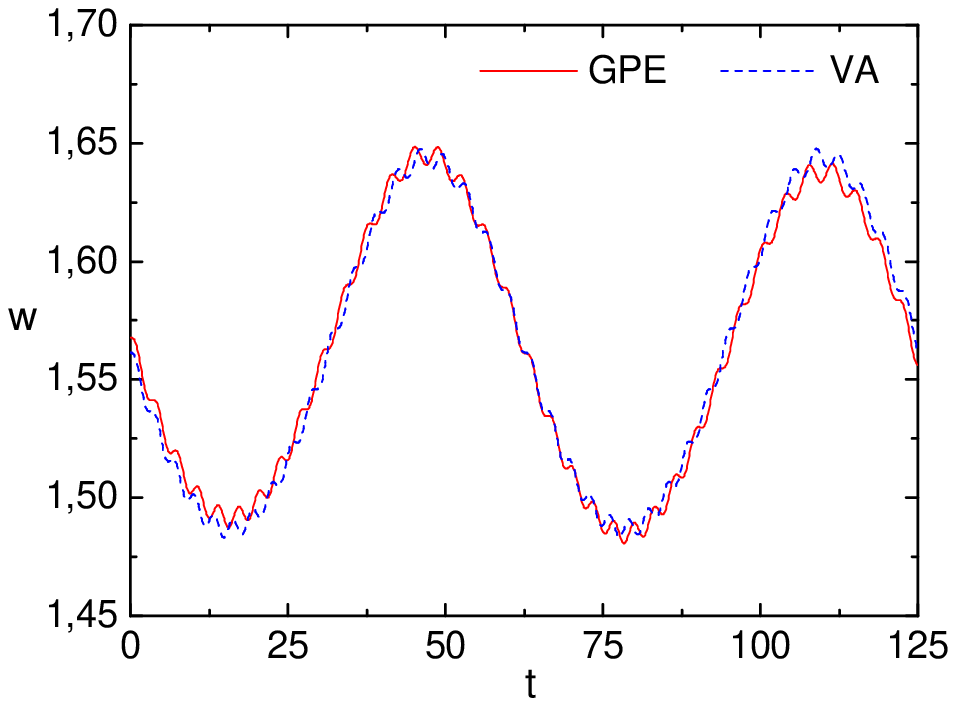}}
\caption{(Color online) Left panel: The potential (\ref{pot}) for
the effective particle model Eq. (\ref{att}) for two sets of
parameters found from stationary version of VA: $\alpha = 100$,
$\beta = 200$, $M=0.155$, $a_0 = 1.41$ and $\alpha = 100$, $\beta
= 400$, $M=0.111$, $a_0 = 2.78$. The minimum of the potential
corresponds to the stationary width ($a_0$) of the flat-top
soliton. Right panel: Oscillation of the amplitude under slowly
varying strength of cubic nonlinearity $\alpha(t)=\alpha_0 [1 +
\delta \sin(\omega t)]$, with $\delta = 0.05$, $\omega = 0.1$,
found from numerical solution of the GPE (\ref{gpe-td}) (red solid
line), and as predicted by VA Eq. (\ref{att}) (blue dashed line).
} \label{fig4}
\end{figure}

\begin{figure}
\centerline{\includegraphics[width=8cm,height=6cm,clip]{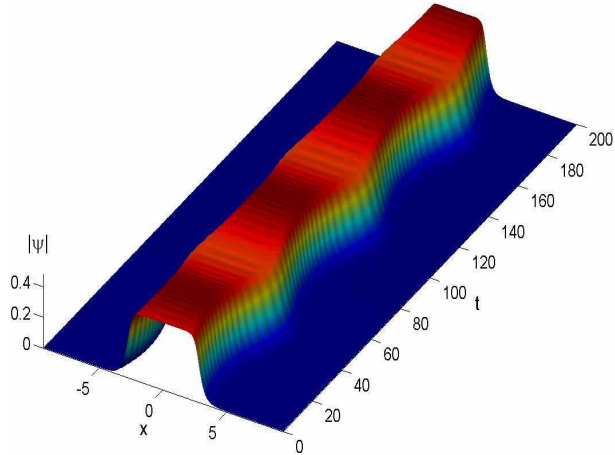}}
\caption{(Color online) Surface plot of a flat-top soliton
evolving under nonlinearity management according to numerical
solution of GPE (\ref{gpe-td}). Parameter settings correspond to
the right panel of Fig. \ref{fig4}. } \label{fig5}
\end{figure}

This equation is analogous to the equation of motion of a unit
mass particle in anharmonic potential $a_{tt} = - \partial
U(a)/\partial a $, with
\begin{equation}\label{pot}
U(a)=\frac{f_1(M)+\beta f_2(M)}{2\,a^2} - \frac{\alpha_0
f_3(M)}{a},
\end{equation}
where $\alpha_0$ is the stationary value of the cubic nonlinear
coefficient. The shape of this potential is depicted in the left
panel of Fig. \ref{fig4}. The minimum of the potential corresponds
to the width of the unperturbed flat-top soliton $a_0$, found from
Eqs.~(\ref{A})-(\ref{M}). Weakly perturbed (chirped) soliton
performs small amplitude oscillations around the minimum of the
potential (\ref{pot}) with a frequency
\begin{equation}\label{omega0}
\omega_0 =[3(f_1(M)+\beta f_2(M))-2a_0 \alpha f_3(M)]^{1/2}/a_0^2.
\end{equation}
Oscillations of the mean square width of a flat-top soliton,
defined as
\begin{equation}
w^2(t) = \int \limits_{-\infty}^{\infty} x^2 |\psi(x,t)|^2 dx =
a^2(t)\, \frac{\Gamma(1+3 M)}{3 \Gamma(1+M)},
\end{equation}
is shown in the right panel of Fig. \ref{fig4}. Apparently, the
dynamics described by the VA Eq. (\ref{att}) agrees quite well
with numerical solution of the GPE~(\ref{gpe-td}). The peculiarity
of the dynamics is that, the fine structure due to fast internal
vibrations of the soliton with a frequency (\ref{omega0}) is
superimposed upon the slow dynamics under nonlinearity management.
For the parameter settings of the right panel of Fig. \ref{fig4},
the frequency of internal vibrations found from numerical solution
of the GPE (\ref{gpe-td}) is $\omega_0 = 1.76$, while the
prediction of the VA according to Eq. (\ref{omega0}) is $\omega_0
= 1.89$. When the frequency of nonlinearity management is close to
this frequency, the dynamics is highly irregular. Surface plot of
the flat-top soliton evolving under slow nonlinearity management
is shown in Fig. \ref{fig5}.

Numerical simulations of the GPE (\ref{gpe-td}) are performed by
means of the split-step fast-Fourier-transform method
\cite{agrawal,press} in a spatial domain of length $L = 8 \pi$
with 2048 modes. The time step was $\Delta t = 0.001$. To control
the numerical results, we monitored the accuracy of normalization
condition (\ref{norm1}), which showed conservation during the time
evolution to precision better than $10^{-3}$ in normalized units.
To prevent re-entering of the linear waves emitted by the
perturbed soliton into the integration domain, absorbers were
installed at domain boundaries. To obtain the stationary profiles
of flat-top solitons in external potentials for Eq. (\ref{gpe1})
we employed the imaginary-time relaxation method for finding
ground states in NLSE-based models \cite{chiofalo}.

\section{Conclusions}

We have developed a variational approach for flat-top solitons,
and employed it for the analysis of static and dynamic properties
of matter-wave solitons which can exist in BEC's with significant
contribution of three-body interactions. The accuracy of the
developed approach has been verified by comparing the predictions
of VA equations with corresponding data from the numerical
solution of the governing GPE, and good agreement is found between
the two results. Although the emphasis was given to BEC
applications, the developed theory is general and may apply to
nonlinear optics phenomena in materials with cubic-quintic
nonlinearity.

\section*{Acknowledgements}

B.B.B. thanks Prof. M. Salerno for valuable comments. A.B. is
grateful to Physical-Technical Institute of the Uzbek Academy of
Sciences for hospitality. This work is supported by the research
grant No.~IIUM/504/RES/G/14/11/02/FRGS0106-29 from the Ministry of
Higher Education of Malaysia, and KFUPM research project No.
IN090008.

\end{document}